\documentclass[a4paper,11pt]{article}
\usepackage{pos}

\def\beqn{\begin{eqnarray}} \def\eeqn{\end{eqnarray}}
\def\beq{\begin{equation}} \def\eeq{\end{equation}}

\title{
\vspace*{-1.5cm}
\begin{minipage}{\textwidth}
{\normalfont\small IFIC/23-42
\hspace{\fill} September 2023
}\\
\end{minipage}\\[60pt]
  Using photon-hadron production to impose restrictions on heavy-hadrons fragmentation functions}

\ShortTitle{Using photon-hadron production to impose restrictions on heavy-hadrons FF}

\author*[a,b]{German F. R. Sborlini}
\author[c]{Roger Hern\'andez-Pinto}
\author[c]{Salvador Ochoa-Oregon}
\author[d]{David F. Renter\'ia-Estrada}

\affiliation[a]{Departamento de F\'isica Fundamental e IUFFyM, Universidad de Salamanca, 37008 Salamanca, Spain.}
\affiliation[b]{Escuela de Ciencias, Ingenier\'ia y Diseño, Universidad Europea de Valencia, Paseo de la Alameda 7, 46010 Valencia, Spain.}
\affiliation[c]{Facultad de Ciencias F\'isico-Matem\'aticas, Universidad Aut\'onoma de Sinaloa, Ciudad Universitaria, CP 80000 Culiac\'an, Mexico.}
\affiliation[d]{Instituto de F\'{\i}sica Corpuscular, Universitat de Val\`{e}ncia -- 
Consejo Superior de Investigaciones Cient\'{\i}ficas, Parc Cient\'{\i}fic, 46980 Paterna, Valencia, Spain.}

\emailAdd{german.sborlini@usal.es}

\abstract{Fragmentation Functions (FF) are universal non-perturbative objects that model hadronization in some general kind of processes. They are mainly extracted from experimental data, hence constraining the parameters of the corresponding fits is crucial for achieving reliable results. As expected, the production of lighter hadrons is favoured w.r.t. heavy ones, thus we would like to exploit the precise knowledge of pion FFs to constraint the shape of kaon (or heavier) FFs. In this talk, we show how imposing specific cuts on photon-hadron production leads to relations between the $u$-started FFs. For doing so, we exploit the reconstruction of momentum fractions in terms of experimentally-accessible quantities and introduce NLO QCD + LO QED corrections to reduce the theoretical uncertainties.}

\FullConference{The European Physical Society Conference on High Energy Physics (EPS-HEP2023)\\
 21-25 August 2023\\
Hamburg, Germany\\}

\begin{document}
\maketitle

\section{Motivation}
\label{sec:Motivation}
A precise phenomenological description of particle production in high-energy collisions is crucial to understand the fundamental constituents of matter. Our current knowledge relies on the Standard Model (SM), a gauge theory that successfully predicts most of the measurements obtained in hadron colliders. However, precision plays a fundamental role, since tiny discrepancies between theory and data could hide new physics signals. 

Solving the complicated equations of SM to extract accurate phenomenological predictions is plagued with challenges and bottlenecks. One of these bottlenecks is related to the description of the hadronization process, in which a bunch of partons (gluons, quarks or other fundamental particles) originate hadrons through non-perturbative interactions. Even if there are models \cite{Metz:2016swz} that could be use to approximate this process, exact solutions are not available. Thus, in order to describe the production of pions, kaons and other hadrons, we rely on Fragmentation Functions (FFs), which are extracted from analysis and fits of experimental data \cite{Hirai:2007cx,Albino:2005mv,Kniehl:2000fe,Kretzer:2000yf,Ritzmann:2014mka,deFlorian:2014xna,deFlorian:2017lwf,Bertone:2017bme,Borsa:2022vvp,AbdulKhalek:2022laj}. Being able to experimentally constraint these FFs is important to reduce the fit errors and obtain more precise predictions.

In Ref. \cite{Ochoa-Oregon:2023ktx}, we propose to use photon-hadron production at colliders to improve the extraction of FFs, specially for heavy hadrons. Since the photon acts as a clean probe of the parton collision, it could help us to reconstruct the parton kinematics with more precision. This knowledge can be then used to relate FFs of different hadrons by comparing the ratios of their production rates. The aim of this article is proving that we can constrain $d^K(z)/d^\pi(z)$ (i.e. the ratio of pion and kaon FF) exploiting the ratio of their cross-sections (i.e. $d\sigma_{\gamma+K}/d\sigma_{\gamma+\pi}$) after imposing proper cuts. 

\section{Reconstructing the parton kinematics}
\label{sec:Reconstructing}
In the context of the parton model, it is worth noticing that momentum fractions are not physical quantities; in other words, we can not directly measure them. Still, they allow us to understand what is going on inside the hadrons and we can relate them to experimentally-accessible quantities. For instance, inspired by the LO kinematics of photon+hadron production, we can define
\beqn
x_{1,REC} &=& \frac{p_T^\gamma}{\sqrt{s_{CM}}} \left(\exp(\eta^\pi)+\exp(\eta^\gamma)\right) \, ,
\label{eq:x1REC}
\\ z_{REC} &=& \frac{p_T^\pi}{p_T^\gamma}\, .
\label{eq:zREC}
\eeqn
Whilst the r.h.s. of Eqs. (\ref{eq:x1REC})-(\ref{eq:zREC}) correspond to a function of $\{p_T^\gamma,\eta^\gamma,p_T^\pi,\eta^\pi\}$, the l.h.s. provide an estimator of the momentum fraction of the parton $a_1$ entering the reaction $a_1+a_2 \to \gamma + a_3$, and the momentum carried by the pion in the hadronization process $a_3 \to \pi$, respectively.

\subsection{Reconstruction at NLO (and beyond)}
\label{ssec:ReconstructingNLO}
The estimators in Eqs. (\ref{eq:x1REC})-(\ref{eq:zREC}) are strictly valid at tree-level because the presence of real radiation associated to higher-order QCD corrections introduces new events with different parton-level kinematics. For instance, at next-to-leading order (NLO) we need to combine events involving 2-to-3 (real radiation) and 2-to-2 (virtual corrections) processes. In order to do so, we first discretize the external (experimentally-accessible) variables $\bar{\cal V}_{\rm Exp}$ and create bins. Then, given a point in the corresponding grid, $p_j=\{p_T^\gamma,\eta^\gamma,\phi^\gamma,p_T^\pi,\eta^\pi,\phi^\pi\}$, we calculate the integrated cross-section $\sigma(p_j)$ and define
\beqn 
(x_1)_j &=& \sum_i \, (x_1)_i \frac{d\sigma}{d x_1}(p_j; (x_1)_i) \, ,
\label{eq:x1_weighted}
\\ (z)_j &=& \sum_i \, (z)_i \frac{d\sigma}{d z}(p_j; (z)_i) \, ,
\label{eq:z_weighted}
\eeqn 
that provide a cross-section-weighted approximation to the partonic momentum fractions $x_1$ and $z$. Once this is done, we need to find the maps
\beq
Y_{REC} := \bar{\cal V}_{\rm Exp} \rightarrow Y_{REAL}\, ,
\eeq
with $Y=\{x_1,x_2,z\}$ and $Y_{REAL}$ given by Eqs. (\ref{eq:x1_weighted})-(\ref{eq:z_weighted}), that will allow us to reconstruct the partonic momentum fractions using only information obtained from experimentally-accessible variables. In Ref. \cite{deFlorian:2010vy}, we proposed the LO-inspired relations Eqs. (\ref{eq:x1REC})-(\ref{eq:zREC}) as an approximated map, and we showed that these formula were highly correlated to the \emph{true} momentum fractions including NLO QCD effects.  

\begin{figure}[h!]
    \centering
    \includegraphics[scale=0.15]{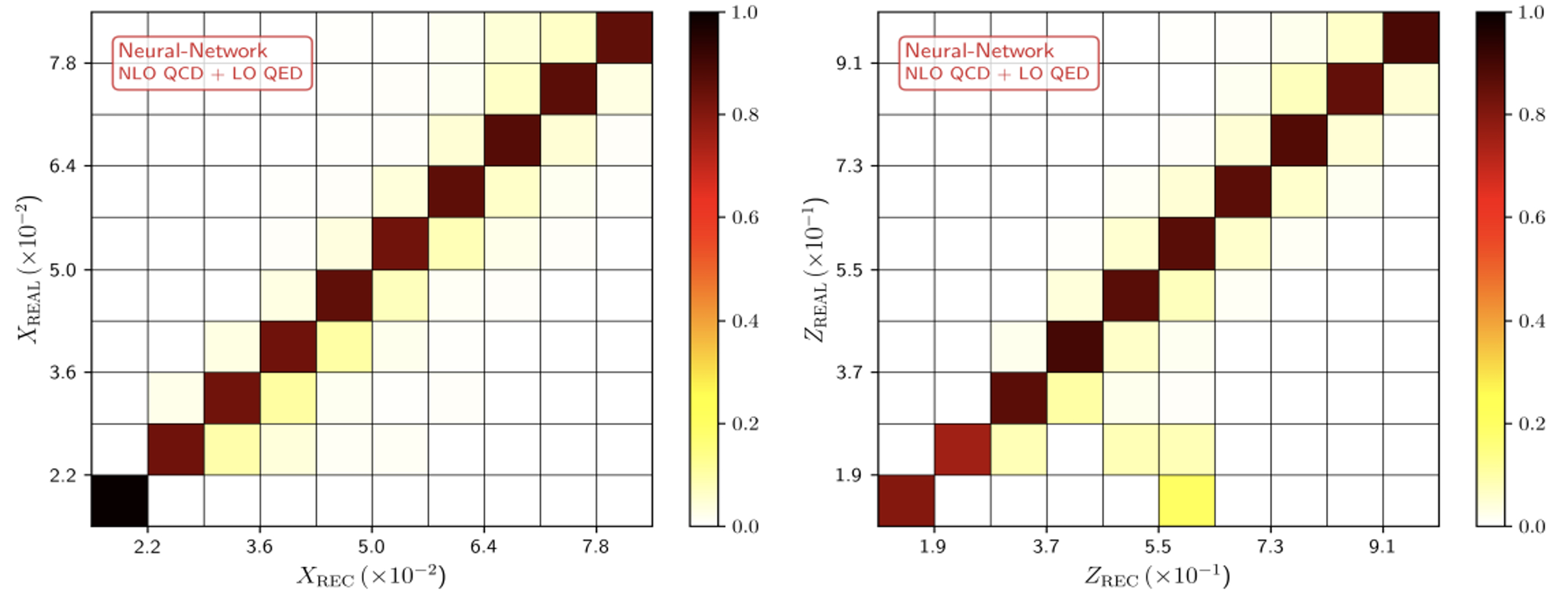}
    \caption{Correlation plots of the \emph{real} vs. reconstructed momentum fractions $x_1$ (left) and $z$ (right), including up to NLO QCD + LO QED effects. We used an optimized neural network based on multilayer perceptrons \cite{Renteria-Estrada:2022scipost}. $\{x_{REAL},z_{REAL}\}$ are the \emph{true} momentum fractions of the events generated by the simulator.}
    \label{fig:NN}
\end{figure} 

Exploiting the recent advances in machine-learning techniques, in Ref. \cite{Renteria-Estrada:2022scipost}, we also used neural networks to find the maps in Eqs. (\ref{eq:x1REC})-(\ref{eq:zREC}). The resulting correlation plots are shown in Fig. \ref{fig:NN}, which exhibit a very accurate reconstruction of the momentum fractions $x_1$ and $z$, with minimal human intervention (no need to define a function basis for the fit).

\section{Constraining Fragmentation Functions}
\label{sec:Constraining}
Once the approximated momentum fractions are described in terms of the external variables, we have access to $z$. We will use this fact to extract information about the FFs. First, we consider the differential cross-section for the process $H_1+H_2\to h_i+\gamma$ as a function of the \emph{real} momentum fraction:
\beqn
\nonumber \frac{d\sigma^{h_i}}{dz_{REAL}} &=&\int dx_1 dx_2 dz \, \sum_{a_1,a_2,a_3} \, d_{a_3}^{h_i}(z) f_{a_1}^{H_1}(x_1) f_{a_1}^{H_1}(x_2) \, d\hat{\sigma}_{a_1 a_2 \to a_3 \gamma}\, \delta(z-z_{REAL})\, 
\\ &=& \sum_{a_1,a_2,a_3} \, d_{a_3}^{h_i}(z_{REAL}) \, g_{a_3}(z_{REAL}) \, ,
\label{eq:factorization}
\eeqn
where $d_{a_3}^{h_i}(z)$ is the FF associated to a parton $a_3$ that hadronizes into $h_i$ carrying a momentum fraction $z$. In order to have the second line of Eq. (\ref{eq:factorization}), we are neglecting the scale dependence, thus having a perfect factorization. Notice that $g_{a_3}(z)$ is independent on the final state hadron $h_i$.

At this point, our aim is clear: exploit Eq. (\ref{eq:factorization}) to find relations among FF for different hadrons. In particular, keeping in mind photon-hadron production, we perform the following approximations:
\begin{enumerate}
    \item Since $z= p_T^{h_i}/p_Y^\gamma=z_{REC}$ is strictly valid at tree-level, we can impose $|\eta|<0.5$ to keep mainly events with Born-level kinematics and use $z \approx z_{REC}$ even when including up to NLO QCD + LO QED corrections \cite{Renteria-Estrada:2021rqp}.
    \item $qg$-initiated channel is roughly 10 times larger than the others, mainly due to gluon PDF enhancement. Having in mind the LO picture, this implies that the dominant production channel at parton level is $q+g \to q + \gamma$, or equivalently that $a_3$ is a quark.
    \item As a consequence of a factor $e_q^2$ in the matrix element, U-channels are 4 times larger than D-channels.
\end{enumerate}
As a consequence of 1, 2 and 3, together with the fact the $u$ is the dominant U-sector quark flavour inside the proton, Eq. (\ref{eq:factorization}) leads to
\beq 
R^{K/\pi}(d\sigma) = \frac{d\sigma^K/dz_{REC}}{d\sigma^\pi/dz_{REC}} \approx \frac{d_u^K(z_{REC})}{d_u^\pi(z_{REC})} = R^{K/\pi}(d_u)\, ,
\label{eq:Ratio1}
\eeq
where we achieve a relation between kaon and pion $u$-started FFs. We considered to initial scenarios to test the validity of this approximation. On one side, we fixed the reference energy scale to $\mu=\bar{Q}=26$ GeV. On the other, we choose the default definition $\mu=(p_T^{h_i}+p_T^\gamma)/2$ which changes event-by-event. In both cases, $R^{K/\pi}(d_u)$ and $R^{K/\pi}(d\sigma)$ exhibit a rather similar shape and they overlap within their corresponding error bands\footnote{As usual, we obtained these bands by varying a factor 2 up and down the renormalization and factorization energy scales. More details are available in Ref. \cite{Ochoa-Oregon:2023ktx}.}. In Fig. \ref{fig:RATIO} (left) we show the results fixing the energy scale to $\mu=\bar{Q}$ without further cuts (except those mentioned in 1).

\subsection{Enhancing different partonic channels}
\label{ssec:Enhancing}
From Eq. (\ref{eq:factorization}), we appreciate that the sum over quark flavours spoils a perfect cancellation of $g_{a_3}$ in the cross-section ratios considered in Eq. (\ref{eq:Ratio1}). Thus, we can impose additional kinematical cuts to enhance even more the contribution of the $u$-quark channel.

\begin{figure}[h!]
    \centering
    \includegraphics[scale=0.105]{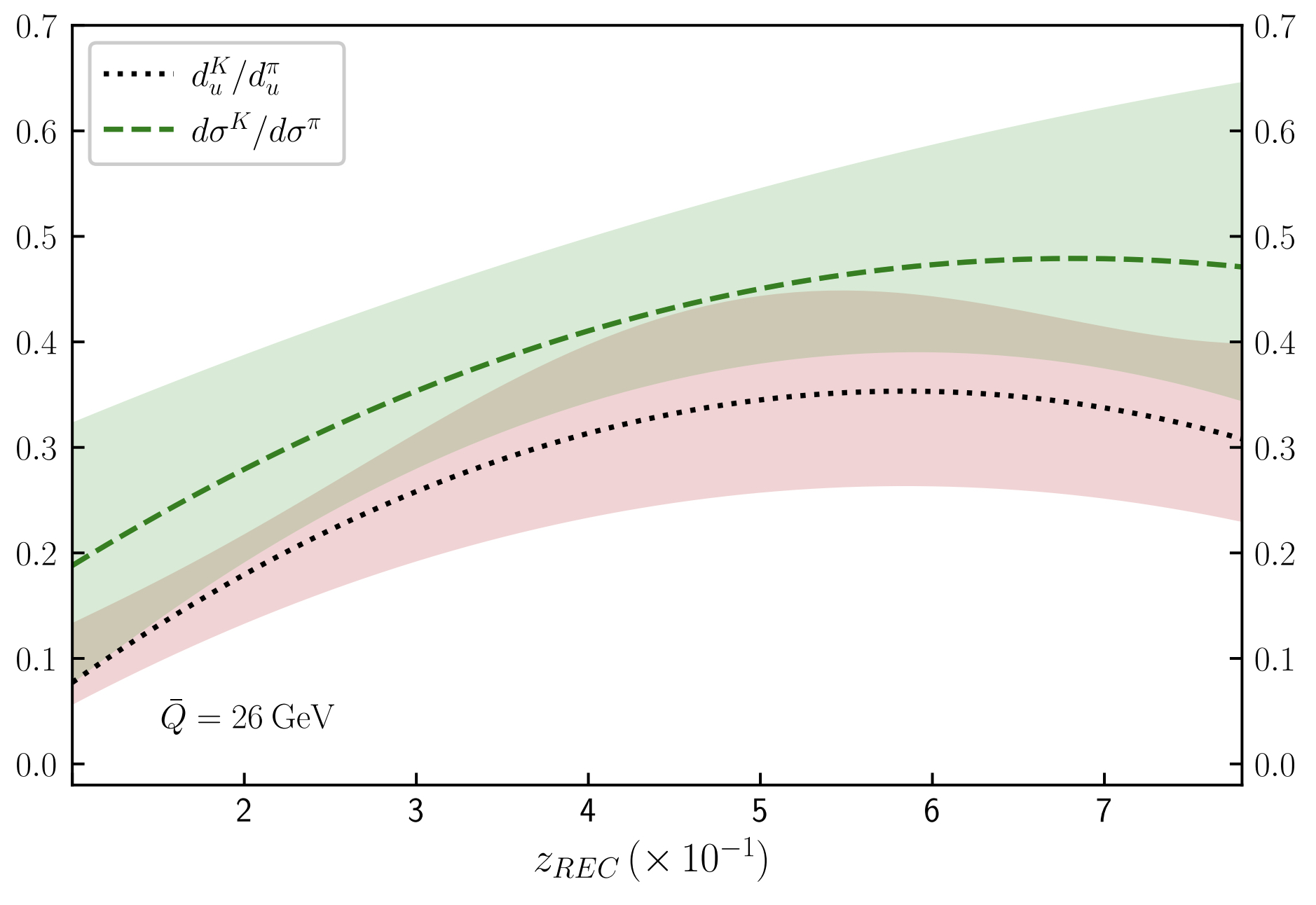} \quad
    \includegraphics[scale=0.105]{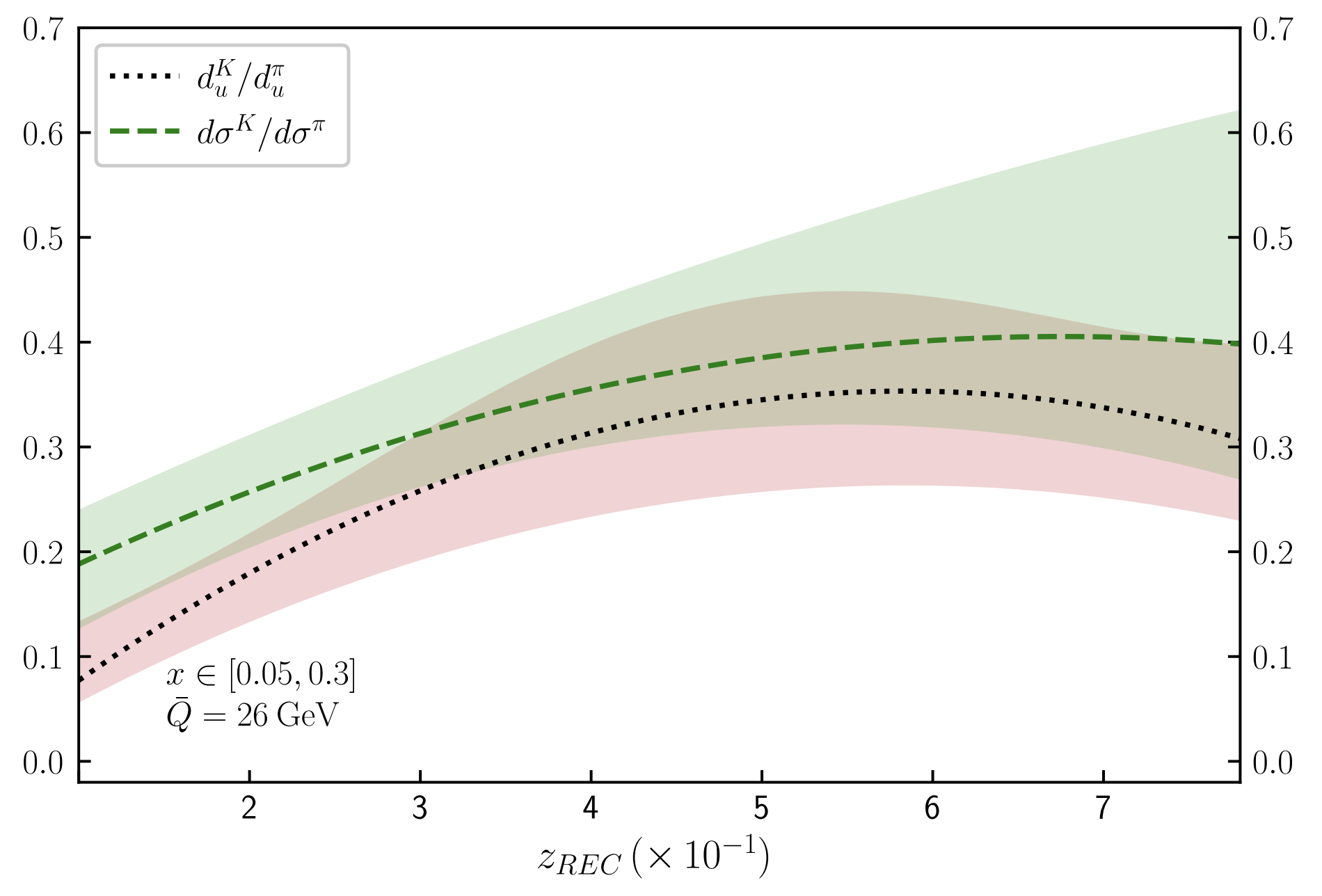}
    \caption{Comparison of the ratios $R^{K/\pi}(d_u)$ (black dashed) and $R^{K/\pi}(d\sigma)$ (green dashed) including up to NLO QCD and LO QED effects. The central energy scale is fixed to $\mu=\bar{Q}$. We show two scenarios: (left) without additional cuts and (right) imposing $0.03\leq \{(x_1)_{REC},(x_2)_{REC}\}\leq0.5$.}
    \label{fig:RATIO}
\end{figure} 

By taking a look to different PDF sets, we notice that $u$ is favoured w.r.t. $d$ for $x \in (0.03,0.5)$. Thus, we used the reconstructed $x$ momentum fractions from Eq. (\ref{eq:x1REC}) and selected those events fulfilling
\beq
0.03\leq \{(x_1)_{REC},(x_2)_{REC}\}\leq0.5 \, .
\eeq
Notice that this cut is totally realistic because $x_{REC}$ is expressed in terms of experimentally-accessible quantities. In Fig. \ref{fig:RATIO} (right) we show the results of this new scenario, where we appreciate that $R^{K/\pi}(d_u)$ and $R^{K/\pi}(d\sigma)$ are much closer. Furthermore, the overlap of their error bands is larger, which indicates that the Eq. (\ref{eq:Ratio1}) is a good approximation.

\section{Conclusions and outlook}
\label{sec:conclusions}
In this article, we motivate the importance of photon-hadron production to access parton-level kinematics, specially when we require the presence of a prompt-photon in the final state. Since this photon acts as a clean probe of the underlying partonic collisions, we can reconstruct the momentum fractions by using experimentally-accessible variables. By using machine-learning tools, we test the validity of the analytic approximations provided in Ref. \cite{deFlorian:2010vy} and show that neural-networks lead to a very accurate reconstruction with minimal human intervention.

Once the momentum fractions are expressed in terms of measurable quantities (such as $p_T$ or $\eta$), we proceed to study cross-section ratios for different hadrons in the final state. By using proper approximations, we manage to relate these ratios to FF ratios. This means that, if we are able to accurately determine FFs for hadron $h_1$, then we can constrain the FF of another hadron $h_2$ by computing $R^{h_2/h_1}(d\sigma)$ as defined in Eq. (\ref{eq:Ratio1}). By imposing optimized kinematical cuts (as the ones described in Sec. \ref{ssec:Enhancing}), we can enhance the contribution of different partonic channels and, in this way, extract more information about the FFs. In the future, we plan to implement machine-learning techniques to optimize the cuts and better constrain FFs for heavy hadrons.

\subsection*{Acknowledgments}
This work is supported by the Spanish Government (Agencia Estatal de Investigaci\'on MCIN /AEI/10.13039/501100011033) Grants No. PID2020-114473GB-I00, PID2022-141910NB-I00; Generalitat Valenciana Grant No. PROMETEO/2021/071. 
G.S. is supported by H2020-MSCA-COFUND USAL4EXCELLENCE-PROOPI-391 project under Grant Agreement No 101034371.
R.H.P. is supported by CONACyT Project No. 320856 ({\em Paradigmas y Controversias de la Ciencia 2022}), {\em Ciencia de Frontera 2021-2042} and {\em Sistema Nacional de Investigadores}.


\bibliographystyle{JHEP}

\begin{thebibliography}{10}

\bibitem{Metz:2016swz}
A.~Metz and A.~Vossen, \emph{{Parton Fragmentation Functions}},
  \href{http://dx.doi.org/10.1016/j.ppnp.2016.08.003}{\emph{Prog. Part. Nucl.
  Phys.} {\bf 91} (2016) 136--202},
  [\href{http://arxiv.org/abs/1607.02521}{{\tt 1607.02521}}].

\bibitem{Hirai:2007cx}
M.~Hirai, S.~Kumano, T.~H. Nagai and K.~Sudoh, \emph{{Determination of
  fragmentation functions and their uncertainties}},
  \href{http://dx.doi.org/10.1103/PhysRevD.75.094009}{\emph{Phys. Rev. D} {\bf
  75} (2007) 094009}, [\href{http://arxiv.org/abs/hep-ph/0702250}{{\tt
  hep-ph/0702250}}].

\bibitem{Albino:2005mv}
S.~Albino, B.~A. Kniehl and G.~Kramer, \emph{{Fragmentation functions for K0(S)
  and Lambda with complete quark flavor separation}},
  \href{http://dx.doi.org/10.1016/j.nuclphysb.2005.11.006}{\emph{Nucl. Phys. B}
  {\bf 734} (2006) 50--61}, [\href{http://arxiv.org/abs/hep-ph/0510173}{{\tt
  hep-ph/0510173}}].

\bibitem{Kniehl:2000fe}
B.~A. Kniehl, G.~Kramer and B.~Potter, \emph{{Fragmentation functions for
  pions, kaons, and protons at next-to-leading order}},
  \href{http://dx.doi.org/10.1016/S0550-3213(00)00303-5}{\emph{Nucl. Phys. B}
  {\bf 582} (2000) 514--536}, [\href{http://arxiv.org/abs/hep-ph/0010289}{{\tt
  hep-ph/0010289}}].

\bibitem{Kretzer:2000yf}
S.~Kretzer, \emph{{Fragmentation functions from flavor inclusive and flavor
  tagged e+ e- annihilations}},
  \href{http://dx.doi.org/10.1103/PhysRevD.62.054001}{\emph{Phys. Rev. D} {\bf
  62} (2000) 054001}, [\href{http://arxiv.org/abs/hep-ph/0003177}{{\tt
  hep-ph/0003177}}].

\bibitem{Ritzmann:2014mka}
M.~Ritzmann and W.~J. Waalewijn, \emph{{Fragmentation in Jets at NNLO}},
  \href{http://dx.doi.org/10.1103/PhysRevD.90.054029}{\emph{Phys. Rev. D} {\bf
  90} (2014) 054029}, [\href{http://arxiv.org/abs/1407.3272}{{\tt 1407.3272}}].

\bibitem{deFlorian:2014xna}
D.~de~Florian, R.~Sassot, M.~Epele, R.~J. Hern\'andez-Pinto and M.~Stratmann,
  \emph{{Parton-to-Pion Fragmentation Reloaded}},
  \href{http://dx.doi.org/10.1103/PhysRevD.91.014035}{\emph{Phys. Rev. D} {\bf
  91} (2015) 014035}, [\href{http://arxiv.org/abs/1410.6027}{{\tt 1410.6027}}].

\bibitem{deFlorian:2017lwf}
D.~de~Florian, M.~Epele, R.~J. Hernandez-Pinto, R.~Sassot and M.~Stratmann,
  \emph{{Parton-to-Kaon Fragmentation Revisited}},
  \href{http://dx.doi.org/10.1103/PhysRevD.95.094019}{\emph{Phys. Rev. D} {\bf
  95} (2017) 094019}, [\href{http://arxiv.org/abs/1702.06353}{{\tt
  1702.06353}}].

\bibitem{Bertone:2017bme}
{\scshape NNPDF} collaboration, V.~Bertone, S.~Carrazza, N.~P. Hartland and
  J.~Rojo, \emph{{Illuminating the photon content of the proton within a global
  PDF analysis}},
  \href{http://dx.doi.org/10.21468/SciPostPhys.5.1.008}{\emph{SciPost Phys.}
  {\bf 5} (2018) 008}, [\href{http://arxiv.org/abs/1712.07053}{{\tt
  1712.07053}}].

\bibitem{Borsa:2022vvp}
I.~Borsa, R.~Sassot, D.~de~Florian, M.~Stratmann and W.~Vogelsang,
  \emph{{Towards a Global QCD Analysis of Fragmentation Functions at
  Next-to-Next-to-Leading Order Accuracy}},
  \href{http://dx.doi.org/10.1103/PhysRevLett.129.012002}{\emph{Phys. Rev.
  Lett.} {\bf 129} (2022) 012002}, [\href{http://arxiv.org/abs/2202.05060}{{\tt
  2202.05060}}].

\bibitem{AbdulKhalek:2022laj}
R.~Abdul~Khalek, V.~Bertone, A.~Khoudli and E.~R. Nocera, \emph{{Pion and kaon
  fragmentation functions at next-to-next-to-leading order}},
  \href{http://dx.doi.org/10.1016/j.physletb.2022.137456}{\emph{Phys. Lett. B}
  {\bf 834} (2022) 137456}, [\href{http://arxiv.org/abs/2204.10331}{{\tt
  2204.10331}}].

\bibitem{Ochoa-Oregon:2023ktx}
S.~A. Ochoa-Oregon, D.~F. Renter\'\i{}a-Estrada, R.~J. Hern\'andez-Pinto and
  G.~F.~R. Sborlini, \emph{{Constraining fragmentation functions through
  hadron-photon production at higher-orders}},
  \href{http://dx.doi.org/10.1103/PhysRevD.107.096002}{\emph{Phys. Rev. D} {\bf
  107} (2023) 096002}, [\href{http://arxiv.org/abs/2303.04965}{{\tt
  2303.04965}}].

\bibitem{deFlorian:2010vy}
D.~de~Florian and G.~F.~R. Sborlini, \emph{{Hadron plus photon production in
  polarized hadronic collisions at next-to-leading order accuracy}},
  \href{http://dx.doi.org/10.1103/PhysRevD.83.074022}{\emph{Phys. Rev. D} {\bf
  83} (2011) 074022}, [\href{http://arxiv.org/abs/1011.0486}{{\tt 1011.0486}}].

\bibitem{Renteria-Estrada:2022scipost}
D.~F. Rentería-Estrada, R.~J. Hernández-Pinto, G.~F.~R. Sborlini and
  P.~Zurita, \emph{{Reconstructing partonic kinematics at colliders with
  machine learning}},
  \href{http://dx.doi.org/10.21468/SciPostPhysCore.5.4.049}{\emph{SciPost Phys.
  Core} {\bf 5} (2022) 049}.

\bibitem{Renteria-Estrada:2021rqp}
D.~F. Renter\'\i{}a-Estrada, R.~J. Hern\'andez-Pinto and G.~F.~R. Sborlini,
  \emph{{Analysis of the Internal Structure of Hadrons Using Direct Photon
  Production}}, \href{http://dx.doi.org/10.3390/sym13060942}{\emph{Symmetry}
  {\bf 13} (2021) 942}, [\href{http://arxiv.org/abs/2104.14663}{{\tt
  2104.14663}}].

\end{thebibliography}

\providecommand{\href}[2]{#2}\begingroup\raggedright\endgroup

\end{document}